\documentclass[aps,prb,showpacs, twocolumn,floatfix,superscriptaddress]{revtex4-1}
\usepackage{graphicx,hyperref,amssymb,amsfonts,amsmath,times,color,soul,epstopdf}  
\usepackage{bbold,bm}
\usepackage{appendix}

\newcommand{\pz}[1]{\sigma^z_{#1}}

\def\bea{\begin{eqnarray}}
\def\eea{\end{eqnarray}}

\def\frac#1#2{{\textstyle{#1 \over #2}}}
\def\yd{^\dagger}
\def\nd{^{\vphantom{\dagger}}}
\def\ns{^{\vphantom{*}}}

\def\CN{{\cal N}}

\def\CW{{\cal W}}
\def\CX{{\cal X}}
\def\CT{{\cal T}}
\def\CZ{{\cal Z}}
\def\CV{{\cal V}}
\def\CR{{\cal R}}
\def\CC{{\cal C}}

\def\HH{{\hat H}}
\def\ie{{\it i.e.\/}}

\def\etal{{\it et al.\/}}
\def\mone{\mathbb{1}}
\def\Tra{\textrm{Tr\,}}

\def\dhat{\hat{\bm d}}
\def\nhat{\hat{\bm n}}
\def\xhat{\hat{\bm x}}

\def\zhat{\hat{\bm z}}
\def\Bsigma{{\bm\sigma}}
\def\Btau{{\bm\tau}}

\def\Bd{{\bm d}}

\def\Rep{{\rm Re}}
\def\Imp{{\rm Im}}
\def\CO{{\cal O}}
\def\pz{\partial}
\def\le{\leqslant}

\def\MR{\mathbb{R}}

\def\dar{\downarrow}

\def\rrangle{\rangle\hskip-2.2pt\rangle}
\def\llangle{\langle\hskip-2.2pt\langle}
\def\vvert{|\hskip -1.2pt |}
\def\sket#1{{\vvert  \, #1 \,  \rrangle}}
\def\sbra#1{{\llangle \,  #1 \, \vvert}}
\def\sbraket#1#2{{\llangle \, #1  \, \vvert  \, #2 \,  \rrangle}}
\def\sexpect#1#2#3{{\llangle \, #1 \, \vvert \,  #2  \, \vvert \, #3 \, \rrangle}}
\def\qket#1{{|  \, #1 \,  \rangle}}

\date{\today}

\newcounter{myparagraphs}

\begin{document}

\title{Fisher zeros and persistent temporal oscillations in non-unitary quantum circuits}

\author{Sankhya Basu}
\affiliation{Physics program and Initiative for the Theoretical Sciences, The Graduate Center, CUNY, New York, NY 10016, USA}
\affiliation{Department of Physics and Astronomy, College of Staten Island, CUNY, Staten Island, NY 10314, USA}
\author{Daniel P. Arovas}
\affiliation{Department of Physics, University of California at San Diego, La Jolla,
California 92093, USA}
\author{Sarang Gopalakrishnan}
\affiliation{Department of Physics, The Pennsylvania State University, University Park, PA 16802, USA}
\affiliation{Physics program and Initiative for the Theoretical Sciences, The Graduate Center, CUNY, New York, NY 10016, USA}
\affiliation{Department of Physics and Astronomy, College of Staten Island, CUNY, Staten Island, NY 10314, USA}
\author{Chris A. Hooley}
\affiliation{SUPA, School of Physics and Astronomy, University of St Andrews, North Haugh, St Andrews, Fife KY16 9SS, United Kingdom}
\author{Vadim Oganesyan}
\affiliation{Physics program and Initiative for the Theoretical Sciences, The Graduate Center, CUNY, New York, NY 10016, USA}
\affiliation{Department of Physics and Astronomy, College of Staten Island, CUNY, Staten Island, NY 10314, USA}

\date{\today}
\begin{abstract}
We present a quantum circuit with measurements and post-selection that exhibits a panoply of space- and/or time-ordered phases, from ferromagnetic order to spin-density waves to time crystals. Unlike the time crystals that have been found in unitary models, those that occur here are \emph{incommensurate} with the drive frequency. The period of the incommensurate time-crystal phase may be tuned by adjusting the circuit parameters.  We demonstrate that the phases of our quantum circuit, including the inherently non-equilibrium dynamical ones, correspond to complex-temperature equilibrium phases of the exactly solvable square-lattice anisotropic Ising model.
\end{abstract}

\maketitle
\section{Introduction}
For a many-body quantum system with Hamiltonian operator $\HH$, there is an evident formal similarity between the unitary time-evolution operator, $e^{-i \HH t/\hbar}$, and the density operator for a thermal equilibrium state, $e^{-\beta \HH}$.  Since the 1950s this has led to very fruitful cross-fertilization between the theory of quantum dynamics and the equilibrium statistical mechanics of quantum systems.  Perhaps the most influential of these is the Matsubara formalism\cite{matsubara1955}, where the thermal density operator is regarded as an evolution operator in imaginary time:\ this allows many of the tools of diagrammatic perturbation theory to be copied more or less directly from the dynamical to the statistical case. 

In recent decades, the development of the theory of open quantum systems has led to a broadening of interest on the dynamical side of the dynamical-statistical correspondence, since interactions between the quantum system of interest and its environment generically induce (effectively) non-unitary evolution.  The quantum circuits in which we shall be mainly interested here exhibit many-body mixed dynamics, with unitary evolution interrupted by projection operations meant to model measurements by a classical environment.  Crucially, the many-body system is allowed to continue evolving after such measurements, and displays a host of novel phenomena due to the tunable interplay of non-unitary measurements and the intrinsic unitary dynamics\cite{nahum,Li2018, Li2019, Chan2019, Cao2019, MPAF,Gullans2019A, Gullans2019B, Zabalo2020, Jian2019, Bao2020, Tang2020, Fan2020, Fisher2020, Lavasani2020, Hsieh2020, ippoliti2020entanglement, Alberton2020, fidkowski2020dynamical}.

There have been parallel broadenings of interest on the statistical side.  Starting from the early 1950s, Lee, Yang, Fisher and others\cite{complextemp,leeyang,yanglee,griffiths69} pioneered the extension of conventional statistical mechanics to the case where
the coupling constants in the Hamiltonian, or even the inverse temperature $\beta$ itself, are considered to be complex quantities.  
This opens up the possibility of points in the complex $\beta$-plane where the partition function vanishes, something which is not
possible for real temperature.  For simple models, such as the isotropic zero-field Ising model on the square lattice, these `Fisher zeros'
occur on contours in the complex $\beta$-plane which cut the real $\beta$-axis at positions corresponding to the critical temperatures
of phase transitions in the model.  If the density of zeros vanishes as the real $\beta$-axis is approached, the transition is continuous; 
if the density remains finite, the transition is first-order.

%
%
\begin{figure}
\begin{center}
\includegraphics[width=0.9\columnwidth]{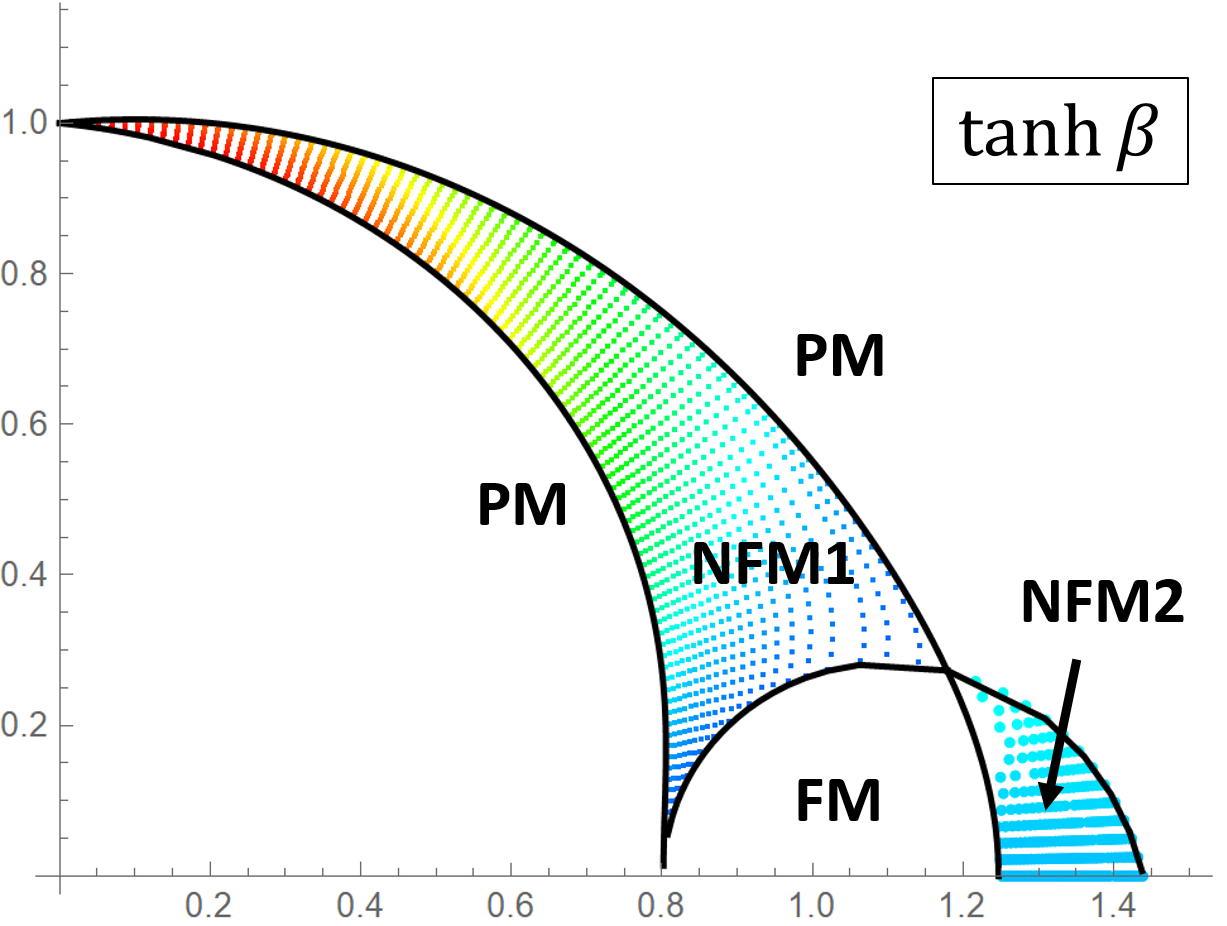}
\end{center}
\caption{Phase diagram showing the various spatially and/or temporally ordered phases of our quantum circuit.  PM (paramagnet):\ short-range order in both space and time.  FM (ferromagnet):\ uniform long-range order in both space and time.  NFM1 (non-ferromagnet 1):\ modulated quasi-long-range order in time (`algebraic time crystal').  NFM2 (non-ferromagnet 2):\ modulated quasi-long-range order in space. The color scheme is used to indicate the evolution of period of oscillations. In particular, it vanishes exactly on the NFM1-FM boundary and reaches maximum at the unitary point $\tanh \beta \to i$ where all damping processes cease and we reach a simple (Floquet) transverse field Ising model spectrum.}
\label{f:phasediag}
\end{figure}
%
%

Complex-coupling approaches to statistical mechanics have tended to be seen as an essentially formal tool.
In light of recent progress in understanding the rich phase structure of \emph{non-unitary} circuits such as those under continuous measurement, it is timely to revisit canonical statistical mechanics models at complex temperature \emph{considered as descriptions of non-unitary evolution}. This is the perspective we adopt here.
%
%
%
%
%
Classical measurements generically introduce randomness in discrete space-time.  By contrast, the canonical Ising model is disorder-free.  The correspondence that we shall demonstrate therefore requires post-selection of measurement outcomes~\cite{lee_chan, biella2020many, gopalakrishnan2020entanglement, nahum2020measurement, jian2021yang}, \ie\  only certain outcomes are allowed to continue evolving and contribute to the eventual disorder-free ensemble of trajectories.  Some of the features of the statistical side of our correspondence were anticipated in prior work by some of us\cite{Beichert2013}, demonstrating the existence of long-range incommensurately modulated correlations underlying Fisher zeros in the thermodynamics of Ising ladders.

We shall show that the dynamics of the corresponding $M$-qubit circuits exhibit long-range correlations that coalesce, in the 
$M\to \infty$ limit, into extended ordered regions (see Fig.~\ref{f:phasediag}).  These can be interpreted as phases of the anisotropic
2D Ising model at complex temperature.  They include relatively conventional short-range ordered `paramagnetic' phases and 
long-range-ordered ferromagnetic and antiferromagnetic phases, but also somewhat peculiar incommensurate critical phases.  These
latter phases exhibit spatially and/or temporally modulated correlators with a dynamically determined modulation period untethered
from the underlying lattice.  At least one of these latter phases bears a phenomenological resemblance to the time crystals recently
discussed in the context of unitary dynamics of isolated many-body localized systems\cite{khemani2019brief}. However, it does not 
fit into the classification presented in that work, since the circuits we consider are non-unitary, and the no-go theorems
\cite{PhysRevLett.114.251603} forbidding time-crystalline order consequently do not apply.

Before turning to our results, we briefly discuss some connections to superficially similar questions discussed in previous literature.
Temporally modulated phases in open quantum systems (\ie\ limit cycles) have been shown to exist in more than two spatial
dimensions~\cite{PhysRevA.91.051601, PhysRevResearch.2.022002}. The non-unitary quantum circuits we consider can be regarded 
as Trotterized non-Hermitian Hamiltonians, which have been extensively explored~\cite{ashida2020non, lee_chan, biella2020many,
gopalakrishnan2020entanglement}. Unlike these works, we keep the Trotter ``time-step'' finite, so the models we consider are 
two-dimensional statistical mechanics models with a transfer matrix that may be contracted either sideways or from top to bottom. 
In addition, many-body entanglement properties have been computed by contracting transfer matrices sideways in a series of recent
works~\cite{PhysRevLett.123.210601, PhysRevB.100.064309, foss2020holographic, garratt2020many, lerose2020influence, 
garratt2020many2, sonner2020characterizing, ippoliti2020postselection, lu2021entanglement, ippoliti2021fractal}, but primarily in
contexts where the dynamics is unitary along one or both directions. In particular, Refs.~\onlinecite{foss2020holographic, 
ippoliti2020postselection, PhysRevLett.118.180601} have proposed experimental protocols to study the non-Hermitian dynamics of 
large systems using spacetime duality. On the statistical mechanics side, Ref.~\onlinecite{GarciaWei} used tensor-network methods
similar to those we use here~\cite{LevinNave} to characterize the thermodynamics of the Yang-Lee model. The present work applies
tools from the complex-temperature statistical mechanics literature~\cite{LevinNave, GarciaWei} to discuss the unexplored physics 
of spatio-temporal correlations in non-unitary quantum circuits. So far these circuits have primarily been studied for their 
entanglement properties; we demonstrate here that even their conventional correlation functions can exhibit striking phenomena 
that would be forbidden by unitarity (in closed systems) or by dimensionality (in open systems~\cite{PhysRevA.91.051601,
PhysRevResearch.2.022002}).

The remainder of the paper is organized as follows.  In Section \ref{s:prelims} we define our quantum circuit by explicitly constructing the local gates necessary to reproduce the complex-temperature statistical mechanics of the Ising model. We also discuss observables of interest, both `thermodynamic' quantities and two-point correlation functions, and construct the transfer matrices that govern the complex-temperature statistical mechanics of interest.  
In Section \ref{s:2D} we study the anisotropic complex-temperature Ising model analytically using fermionization and also numerically using a tensor-network coarse-graining scheme.  The fermionization treatment is restricted to the case without an externally applied magnetic field, but the tensor-network coarse-graining approach allows us to identify additional transitions as a function of field strength.  We conclude with a summary and a discussion of open problems.

\section{Quantum circuits, observables, and transfer matrices}
\label{s:prelims}

%
%
%
%

\begin{figure}[!b]
\begin{center}
\includegraphics[width=0.95\columnwidth]{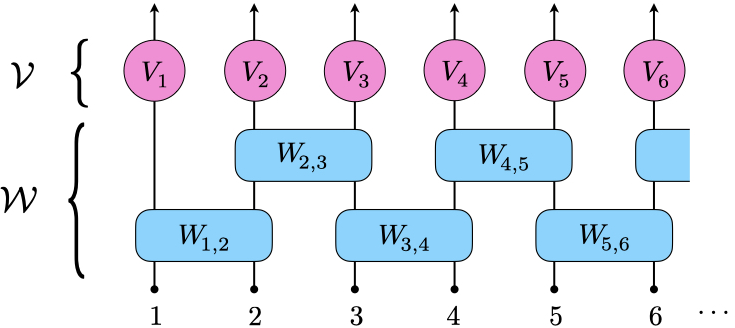}
\end{center}
\caption{Sketch of the quantum circuit defined in eqn. (\ref{qcir}).}
\label{Fcircuit}
\end{figure}

\subsection{Quantum circuit formalism}\label{secQCF}
A quantum circuit consists of a set of qubits, which we may label $\{1,\ldots,M\}$, successively subjected to operations
in the form of quantum gates.  In the scenario investigated here, the qubits are Ising spins and the gates are of two types:
two-qubit gates $W\nd_{m,m+1}=\exp(JZ\nd_m Z\nd_{m+1})$ and single qubit gates
 $V\nd_m=A\exp(\gamma X\nd_m)$, where $\{X\nd_m,Y\nd_m,Z\nd_m\}$ are the Pauli spin matrices for qubit $m$.  Defining
\begin{equation}
\begin{split}
\CW&=W\nd_{1,2}\cdots W\nd_{M-1,M}=\exp\!\bigg(J\sum_{m=1}^{M-1}Z\nd_m\otimes Z\nd_{m+1}\bigg)\\
\CV&=V\nd_1 V\nd_2\cdots V\nd_M=A^M\exp\!\bigg(\gamma\sum_{m=1}^M X\nd_m\bigg)\quad,
\label{qcir}
\end{split}
\end{equation}
the single-step evolution operator $\CT$ for each temporal slice of the circuit is $\CT=\CV\CW$, and the evolution operator
for the full circuit is $\CT^L$, where $L$ is the total number of temporal slices.  Note that our qubits are arranged in a finite 
chain {\it vis-{\`a}-vis\/} the operator $\CW$, rather than a ring with periodic boundary conditions.  This is a common context 
for models of quantum circuits; what is new here is that the parameters $J$, $\gamma$, and $A$ are all allowed to be complex, 
hence the individual quantum gates are not in general unitary.  Acting on a density matrix $\rho\ns_j$, the result  of each 
temporal slice of the circuit is the update $\rho\ns_{j+1}=\CT\rho\ns_j\CT\yd$.   Requiring that $\Tra\,\rho\ns_j$ be preserved
then imposes a relationship among the complex parameters $\{J,\gamma,A\}$.

\subsection{Single-qubit transfer matrix}
At each site, our single-qubit transfer matrix $V$ is a product of a unitary $U=\exp(i\omega \dhat\cdot\Bsigma)$ and a 
POVM (positive operator-valued measure),
\begin{equation}
P(\nhat,\phi\,|\,\alpha)=\frac{1}{\sqrt{2}}\big(\cos\phi\ \mone + \alpha\sin\phi\,\nhat\cdot\Bsigma\big)\quad,
\end{equation}
where both $\dhat$ and $\nhat$ are unit vectors on $\textsf{S}^2$, and where $\alpha=\pm 1$ is the specified measurement 
outcome.  The fact that $\sum_\alpha P\yd(\nhat,\phi\,|\,\alpha)\, P(\nhat,\phi\,|\,\alpha)=\mone$ is what makes 
$P(\nhat,\phi\,|\,\alpha)$ a POVM; the fact that there is one such operator for each measurement outcome $\alpha$ means 
that the measurement is `efficient'.  We choose $\dhat=\nhat=\xhat$, and we write
\begin{equation}
V(\alpha)=e^{i\omega X}\cdot\frac{1}{\sqrt{2}}\big(\cos\phi\ \mone + \alpha\sin\phi\,X\big)\equiv A\, e^{\alpha\gamma X}
\end{equation}
where $A=\sqrt{\cos(2\phi)/2}$ and
\begin{equation}
\Rep\,\gamma=\tanh^{-1}\tan\phi\quad,\quad\Imp\,\gamma=\omega\quad.
\end{equation}
We define the operator
\begin{equation}
\CV(\alpha\ns_1,\ldots,\alpha\ns_M)\equiv V\ns_1(\alpha\ns_1)\cdots V\ns_M(\alpha\ns_M)\quad.
\end{equation}
Below we shall post-select $\alpha\ns_m=1$ for all sites, hence
\begin{equation}
\CV\equiv\CV(1,1,\ldots,1)=A^M\exp\!\bigg(\gamma\sum_{m=1}^M X\ns_m\bigg)\quad.
\end{equation}

The single site transfer matrix $V=A\exp(\gamma X)$ is identical to that of the one-dimensional classical Ising model
$H=-J_x\sum_{j=1}^L\sigma\ns_j\sigma\,\ns_{j+1}$\,:
\begin{equation}
V\ns_{\sigma\sigma'}=e^{\beta J_x\sigma\sigma'}=e^{\beta J_x}\big(\mone + e^{-2\beta J_x}X\big)\ns_{\sigma\sigma'}
\end{equation}
with $\tanh\gamma=e^{-2\beta J_x}$ and $A=\sqrt{\coth\gamma-\tanh\gamma}$\,.  Throughout the remainder of
this paper we shall set $J_x\equiv 1$.  Assuming periodicity in this (temporal)
direction, the classical partition function is $\CZ\ns_L(\beta)=\Tra V^L=\cosh^L\!\!\beta+\sinh^L\!\!\beta$ and the
condition $\CZ\ns_L(\beta)=0$ requires $\tanh\beta=e^{(2\ell+1)\pi i/L}$, occurring at $L$ equally spaced values 
$\ell\in\{0,\ldots,L-1\}$ around the circle $|\tanh\beta\,|=1$ in the complex $\tanh\beta$ plane, as noted by Beichert \etal
\cite{Beichert2013}  Interleaved with these Fisher zeros are $L$ points $\tanh\beta=e^{2\pi i\ell/L}$ where the correlation function
\begin{equation}
\begin{split}
C(r;\beta)=\langle\sigma\ns_j\sigma\ns_{j+r}\rangle &=\Tra (Z \, V^r Z \, V^{L-r})\\
&={\tanh^r \!\beta + \tanh^{L-r}\!\beta\over 1+\tanh^L\!\!\beta}
\end{split}
\end{equation}
is long ranged, with $C(r,\beta)=\cos(2\pi \ell r/L)$ (restricting $0\le r \le L$), corresponding to a wavevector
$q={\rm arg}\tanh\beta=2\pi\ell/L$.  In the thermodynamic limit $L\to\infty$, the Fisher zeros coalesce into a branch
cut along the unit circle, with the free energy exhibiting a simple first-order-like cusp nonanalyticity across the cut.

Along the contour of Fisher zeros $\tanh \beta = e^{i \theta}$, with $\theta$ real.  It follows that 
$e^{-2 \beta}=\tanh\gamma=-i \tan(\theta/2)$.  Thus the entire unit circle in the complex $\tanh \beta$ plane 
corresponds to simple unitary stroboscopic precession, \ie\ coherent spin-flipping.  The value $\theta=0$ corresponds 
to a perfectly static spin with no flipping and perfect persistence, analogous to the ferromagnetic ground state of the corresponding
statistical mechanics problem.  The value $\theta=\pi$ corresponds to a complete spin flip with no identity component and 
no persistence, analogous to the antiferromagnetic ground state.

Values of $\tanh \beta$ that do not lie on the unit circle are associated with circuits that include measurement.  Such circuits in the single-qubit case generically exhibit exponentially decaying temporal correlations, just as the analogous complex-temperature Ising models exhibit exponentially decaying spatial correlations.  However, as we shall see, the multi-qubit case is richer:\ when $M>1$ we shall find `decoherence-free subspaces' along lines in the complex-temperature plane, though fully unitary evolution occurs only at isolated points.

\subsection{Two-qubit transfer matrix}
As noted above, our single step evolution operator is a product of single qubit and two qubit terms.
For $M>1$ we introduce the two-qubit transfer matrix $W\ns_{m,m+1}$, which
can be expressed as conditioning a symplectic operation $\exp(\eta J Z\ns_m)$ on the
POVM $P\ns_{m+1}(\zhat,\frac{\pi}{4}\,|\,\eta)$, where $\eta=\pm 1$, for each qubit $m\in\{1,\ldots,L-1\}$.  Explicitly, 
we have
\begin{equation}
\begin{split}
W\ns_{m,m+1}&\equiv e^{JZ\ns_m}\otimes P\ns_{m+1}(\zhat,\frac{\pi}{4}\,|\,1) \\
&\hskip 0.6in + e^{-JZ\ns_m}\otimes P\ns_{m+1}(\zhat,\frac{\pi}{4}\,|\,-1)\\
&=\exp(JZ\ns_m\otimes Z\ns_{m+1})\quad,
\end{split}
\end{equation}
and $\CW\equiv\prod_{m=1}^{M-1} W\ns_{m,m+1}$ as given in eqn. (\ref{qcir}). 


\subsection{Mapping to complex-temperature statistical mechanics}
The correspondence between statistical mechanics in $d+1$ dimensions and quantum mechanics in $d$ dimensions is well known~\cite{PhysRevB.14.1165}.  Traditionally it requires fine-tuning to a critical point of some sort (the so-called 
$\tau$-continuum limit) to enable passing to continuum time where the correspondence is most powerful.  More generally, 
however, the correspondence is to a discrete-time `kicked' quantum evolution, of the type exhibited by our quantum circuit.

By construction, the one-step evolution operator of our quantum circuit resembles the transfer matrix of a statistical mechanical 
system:\ specifically, an anisotropic 2D Ising model.  That model is characterized by its couplings in the $x$- and $y$-directions, 
$J_x$ and $J_y$ respectively, and by its inverse temperature, $\beta$.  These are functions of the quantum circuit parameters 
$\{J,\gamma,A\}$.  In what follows, we shall consider the subspace of circuit parameters for which the inverse temperature 
$\beta$ is complex, while the coupling constants $J_x$ and $J_y$ are real.

One important question is which direction in our two-dimensional statistical model will be identified as the time direction in the
 quantum circuit.  In isotropic models one typically chooses a diagonal direction; here, by contrast, we choose the $x$-axis of our 
 2D anisotropic Ising model -- the one with the stronger coupling ($J_x > J_y$) -- to correspond to the time direction of our 
 quantum circuit.

\subsection{The $M$-qubit transfer matrix}
For complex $\{J,\gamma,A\}$, the transfer matrix $\CT=\CV\CW$, which is of dimension $2^M\times 2^M$, is in general
not normal, \ie\ it does not in general commute with its Hermitian conjugate.  Nevertheless, any non-normal complex matrix
$\CV$ can be brought to Jordan canonical form by a similarity transformation $\CT'=\CR^{-1} \CT\CR$, where $\CR$ is 
invertible.  If we assume there are no Jordan blocks, then $\CT$ may be decomposed in terms of its
eigenvalues and its left and right eigenvectors, {\it viz.}
\begin{equation}
\CT=\sum_{a=0}^{2^M-1} \lambda\ns_a\,\sket{R\ns_a}\sbra{L\ns_a}\quad,
\end{equation}
where $\sbraket{L\ns_a}{R\ns_b}=\delta\ns_{ab}$\,, and where there is no complex conjugation implied in the bra
vector $\sbra{L\ns_a}$ with a doubled bracket.  The eigenvalues $\{\lambda\ns_a\}$ are in general complex. If we order 
the eigenvalues such that  $|\lambda\ns_a| > |\lambda\ns_{a+1}|$ for all $a$, then assuming the largest eigenvalue $\lambda\ns_0$ is nondegenerate,  after a sufficiently large number of iterations $s$ we have
\begin{equation}
\CT^s=\lambda_0^s\,\sket{R\ns_0}\sbra{L\ns_0} + \CO\big(|\lambda\ns_1/\lambda\ns_0|^s\big)\quad.
\end{equation}
It is convenient to here and henceforth implement a similarity transformation and redefine 
$\CT\equiv\CV^{1/2}\,\CW\,\CV^{1/2}$, which is manifestly symmetric: $\CT=\CT^\top$.  The corresponding right and left
eigenvectors of $\CT$ are then mutual transposes, with no complex conjugation, which we write as $\sket{\Psi\ns_a}$ and 
$\sbra{\Psi\ns_a}$\,, respectively.

We consider two natural correlation functions which may be used to characterize the properties of the circuit.  The first is the
quantum two time correlator,
\begin{equation}
C(s;i,j)={\Tra \big[ Z\ns_i\CT^s Z\ns_j\,\rho\ns_0\,(\CT\yd)^s\big]\over
\Tra\big[\CT^s\,\rho\ns_0\,(\CT\yd)^s\big]}\quad.
\end{equation}
With $\rho\ns_0=\mone$, we have
\begin{align}
C(s;i,j)&={\sexpect{\Psi^*_0}{Z\ns_i}{\Psi\ns_0}\sexpect{\Psi\ns_0}{Z\ns_j}{\Psi^*_0}\over
\big|\sbraket{\Psi^*_0}{\Psi\ns_0}\big|^2}\ +\\
&\hskip -0.2in 2\,\Rep\,\Bigg\{\!\bigg({\lambda\ns_1\over\lambda\ns_0}\bigg)^{\!\!s} 
\>{\sexpect{\Psi^*_0}{Z\ns_i}{\Psi\ns_1}\sexpect{\Psi\ns_1}{Z\ns_j}{\Psi^*_0}\over
\big|\sbraket{\Psi^*_0}{\Psi\ns_0}\big|^2}\Bigg\}+ \ldots \nonumber
\end{align}

The second is the statistical correlator,
\begin{align}
C\ns_L(s;i,j)&={\Tra\big[Z\ns_i\CT^s Z\ns_j\CT^{L-s}\big]\over\Tra\big[\CT^L\big]}\\
&=\sexpect{\Psi\ns_0}{Z\ns_i}{\Psi\ns_0}\sexpect{\Psi\ns_0}{Z\ns_j}{\Psi\ns_0}\ +\nonumber\\
&\quad\bigg({\lambda\ns_1\over\lambda\ns_0}\bigg)^{\!\!s}\sexpect{\Psi\ns_0}{Z\ns_i}{\Psi\ns_1}
\sexpect{\Psi\ns_1}{Z\ns_j}{\Psi\ns_0}\ + \nonumber \\
&\quad\bigg({\lambda\ns_1\over\lambda\ns_0}\bigg)^{\!\!L-s}\sexpect{\Psi\ns_1}{Z\ns_i}{\Psi\ns_0}
\sexpect{\Psi\ns_0}{Z\ns_j}{\Psi\ns_1}+ \ldots\nonumber
\end{align}

Let $\CX=\prod_{j=1}^M X\ns_j$.  Since $[\CT,\CX]=0$, assuming $\sket{\Psi\ns_0}$ is nondegenerate,
$\CX\,\sket{\Psi\ns_0}=\pm\,\sket{\Psi\ns_0}$.  Then
\begin{equation}
\sexpect{\Psi\ns_0}{Z\ns_j}{\Psi\ns_0}=\sexpect{\Psi\ns_0}{\CX Z\ns_j\CX}{\Psi\ns_0}=
-\sexpect{\Psi\ns_0}{Z\ns_j}{\Psi\ns_0}
\end{equation}
and thus the $s$-independent terms in the above two correlators both vanish.  We then have that both $C(s;i,j)$ and 
$C\ns_L(s;i,j)$ decay exponentially in the time direction with a correlation time $\tau=1/\ln|\lambda\ns_0/\lambda\ns_1|$ 
and a frequency $\omega=\textrm{arg}(\lambda\ns_1/\lambda\ns_0)$ which is generally incommensurate (\ie\ irrational).  
When the spectral gap collapses, both correlation functions become long-ranged.

At short times we do not expect them to agree, {\it e.g.\/}, for {\it unitary\/} circuits, quantum correlators obey rigid Lieb-Robinson bounds with 
strictly vanishing correlators outside the light cone, while statistical correlators are small but finite for spacelike separations at short times. 

It will also be useful to extend some of our expressions from real-temperature thermodynamics to complex inverse temperature, 
$\beta$.  We examine the modulus the partition function and define the free energy density accordingly 
$f \equiv \log |Z|/ML\beta$, where $M\times L$ is the total number of spins, followed by the internal energy density and 
the specific heat capacity
\begin{equation}
u\equiv \left\vert {\pz f\over\pz\beta} \right\vert\qquad,\qquad c\equiv \left\vert {\pz u\over \pz \beta} \right\vert\quad.
\end{equation}
We typically plot our results not in the complex $\beta$-plane, but rather in the complex $\tanh \beta$ plane, which we shall refer to simply as the `complex temperature plane'.  

\section{Large-$M$ limit:\ the anisotropic 2D Ising model}
\label{s:2D}
In this Section we consider our quantum circuit in the limit of a large number of qubits, $M \gg 1$.  In the $M \to \infty$ limit, the dynamical correlation functions of the circuit may be written in terms of the statistical correlations of an anisotropic 2D Ising model at complex temperature.  We analyze this model using three complementary methods:\ analytic continuation of the Onsager solution; numerical evaluation of a tensor-network representation of the partition and correlation functions; and exact fermionization of the zero-field problem using the Jordan-Wigner transformation.

\subsection{Thermodynamics from the Onsager solution}
\label{s:onsager}
In the isotropic Ising model, the zeros of the partition function lie on linear contours in the complex-temperature plane.  In such a case, provided that the linear density of zeros reaches a finite value in the thermodynamic limit, we expect a simple slope discontinuity in the free energy, as already noted by Fisher \cite{Fisher1965}.  The anisotropic Ising model was examined similarly\cite{}; however, in that case we observe a far more complicated situation with patterns of zeros that appear to occupy extended regions in the complex-temperature plane.  This makes the expected behavior of the free energy less clear.

\begin{figure}
\begin{center}
\includegraphics[width=0.95\columnwidth]{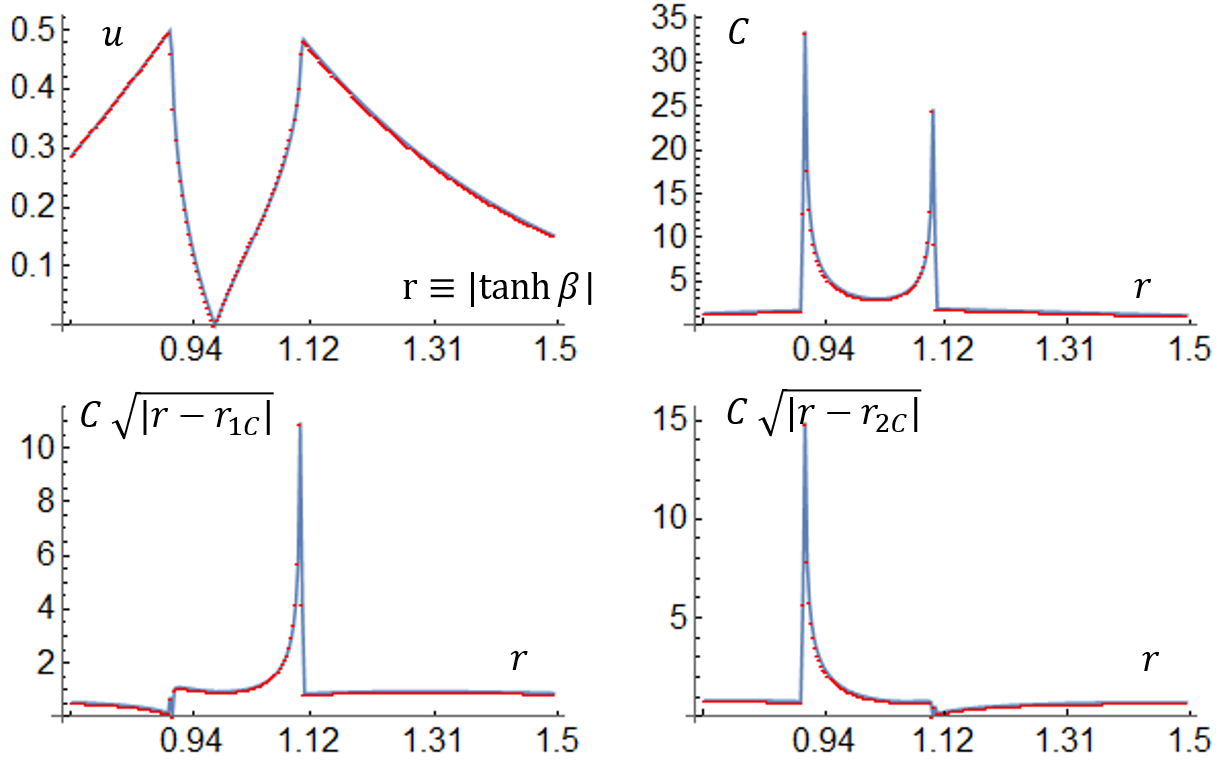}
\end{center}
\caption{Top left: The internal energy as a function of distance along a radial line at angle $2 \pi/9$ to the positive real axis in the complex $\tanh \beta$ plane.  The two transitions are those into and out of the `NFM1' phase (see Fig.~\ref{f:phasediag}).  Top right: The specific heat capacity along the same contour (see text for the precise definition of `specific heat capacity' at complex temperature).  Bottom panels: The same specific heat graph, but multiplied by the specified factor, demonstrating that the two singularities are one-sided square-root singularities.}
\label{f:energyspecheat}
\end{figure}

We note at this point that Fisher's observation and the majority of others that have followed it are in fact based on a portion of Onsager's result; as demonstrated in appendix \ref{app:ladders}, it is manifestly incorrect for the case of finite-width Ising ladders (corresponding to circuits with a finite number of qubits).  Nevertheless, one might anticipate that the approximation remains asymptotically exact for 2D bulk (intensive) quantities. The Onsager expression for the real part of the (dimensionless) free energy per spin of the anisotropic model is
\begin{equation}
\begin{split}
\beta f & =  \ln 2 + {1\over 2} \int\limits_{-\pi}^{\pi} \!{dk_x\over 2\pi} \int\limits_{-\pi}^\pi \!{dk_y\over 2\pi}
\,\ln \big| \cosh j_x \cosh j_y \\
&  \hskip 0.6in  -\,\sinh j_x \cos k_x-\sinh j_y \cos k_y \big|\quad,
\label{e:ZOnsager}
\end{split}
\end{equation}
where $j_{x,y} \equiv 2\beta J_{x,y}$, where $J_x\equiv 1$ and $J\ns_y\equiv J$.
For the case $J_y=0.1$ and complex $\beta$ we have evaluated this for the infinite system numerically.  
Fig.~\ref{f:phasediag} was obtained by taking numerous cuts through the complex temperature plane.  One particular cut that is especially revealing is a radial cut away from the real-temperature axis (Fig.~\ref{f:energyspecheat}) that clearly displays the continuous nature of the PM-NFM1 transition.

%
%

\subsection{Correlations from tensor-network renormalization}
\label{s:trg}
We would like to characterize the different phases that appear in Fig.~\ref{f:phasediag}, especially the NFM1 and NFM2 phases that are not simple continuations of real-temperature phases.  For this we need to know the spin-spin correlation functions in both the $x$ (time) and $y$ (qubit array) directions.  Our most general method for determining these, which has the additional advantage of allowing the inclusion of a longitudinal magnetic field, is via a renormalization group algorithm based on tensor networks - in particular, the Tensor Renormalization Group (TRG)~\cite{LevinNave}, a method that involves representing classical partition functions as tensor networks and coarse-graining these tensor networks numerically.

Most of our results are obtained with bond dimensions up to $50$, and we use relatively modest convergence goals which we check throughout, e.g.\ that the free energy density is converged to $\sim 0.001$.
In the $J_y \to 0$ limit, our system is a set of uncoupled Ising chains; we know that, in this limit, the entire complex-temperature plane is paramagnetic with the exception of the unit circle $|\tanh \beta|=1$.  We therefore expect that, for $J_y \ll 1$, correlated non-paramagnetic phases will be concentrated near the unit circle.

\begin{figure}[h!]
\begin{center}
\includegraphics[width=0.85\columnwidth]{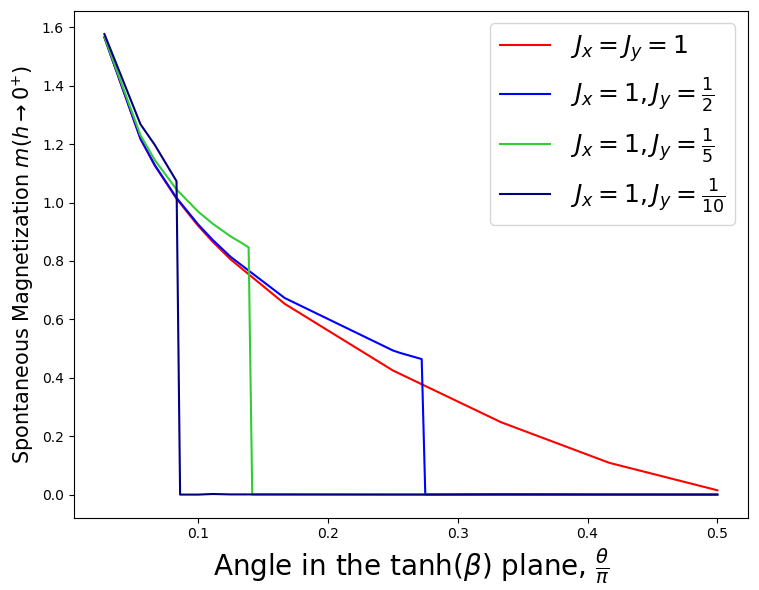}
\end{center}
\caption{The spontaneous magnetization of the anisotropic 2D Ising model as a function of angle from the real-temperature axis on the contour $\vert {\tanh \beta} \vert = 1$, for four different values of the interchain coupling $J_y$.  The jump corresponds to the FM-NFM1 transition, shown for the $J_y = 0.1$ case in Fig.~\ref{f:phasediag}.}
\label{f:magvsanis}
\end{figure}

We have used finite-field TRG to establish the disappearance of uniform ferromagnetic order as we traverse the unit circle $|\tanh \beta| =1$ counterclockwise from the real-temperature line, for several values of the coupling $J_y$ --- see Fig.~\ref{f:magvsanis}.  It is clear that the FM phase is progressively suppressed as the interchain coupling is reduced.  We interpret this as the gradual reversion to the incommensurately modulated order seen on the unit circle in the case of decoupled chains.  This perspective already suggests that the NFM1 phase exhibits some form of long-range incommensurate order; we show below that that is essentially true.
%
%
%
%

Additional work is required to use TRG to compute correlation functions.  As the method is multiplicative, \ie\ distances are reduced by a factor of 2 per iteration, evaluating the correlation function at separations $2^n$ is relatively easy.  These are useful in cases for which we expect simple power-law decays.  Here, however, we are interested in modulated correlators, which requires careful renormalization at short distances.

\begin{figure*}
\begin{center}
\includegraphics[width=0.68\columnwidth]{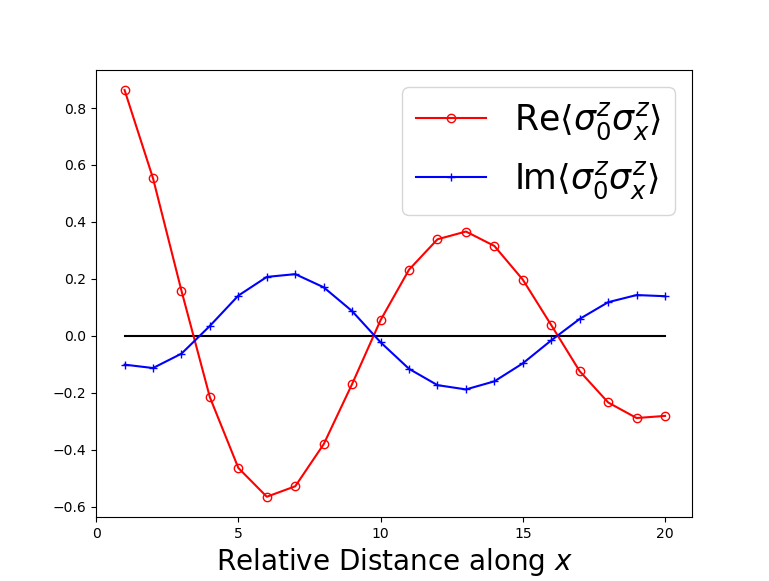}
\includegraphics[width=0.68\columnwidth]{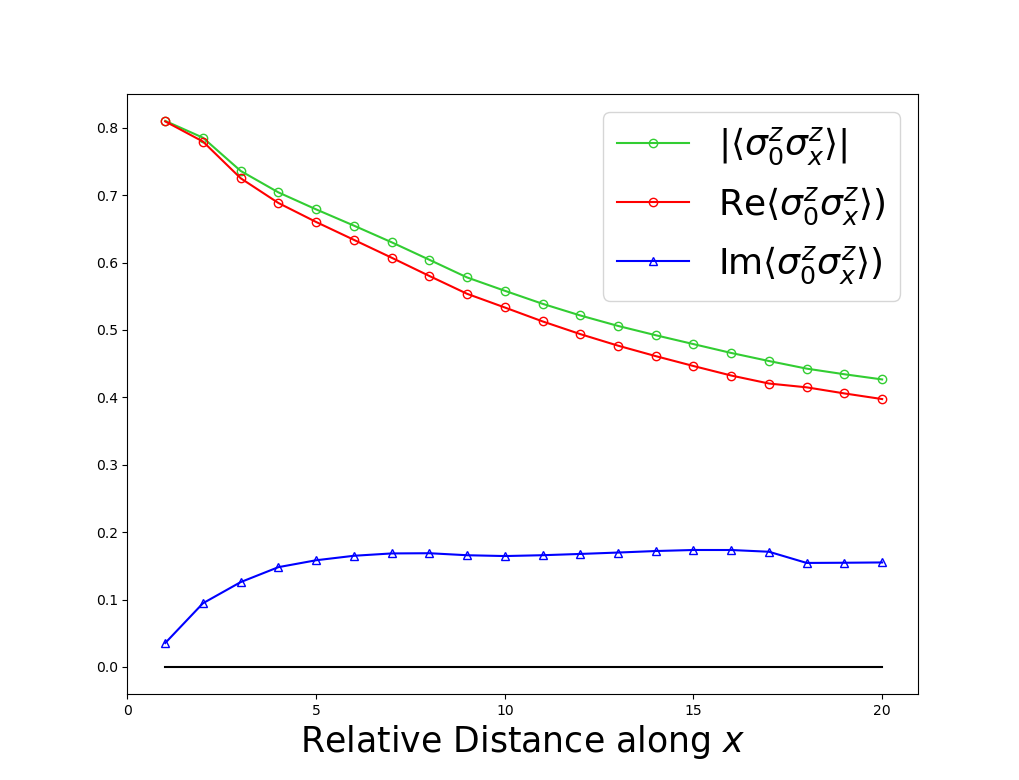}
\includegraphics[width=0.68\columnwidth]{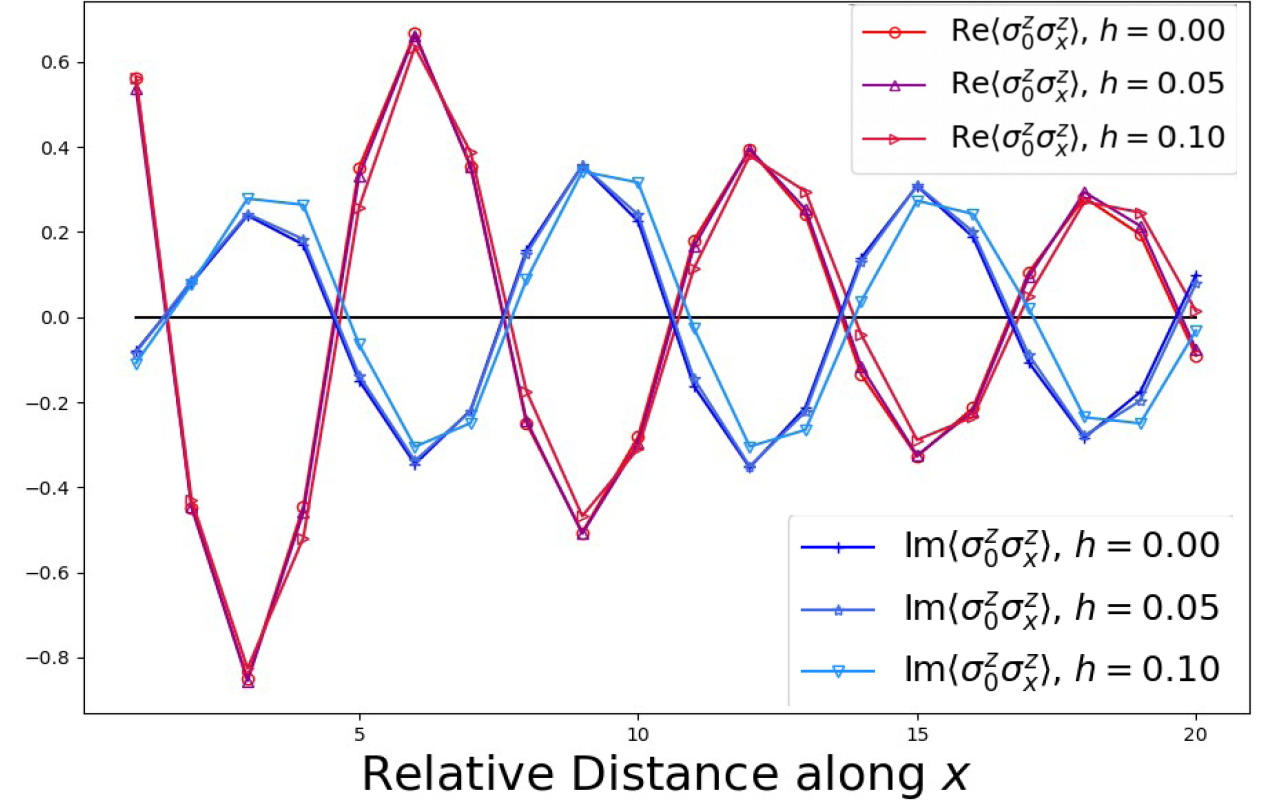}
\\
\includegraphics[width=0.68\columnwidth]{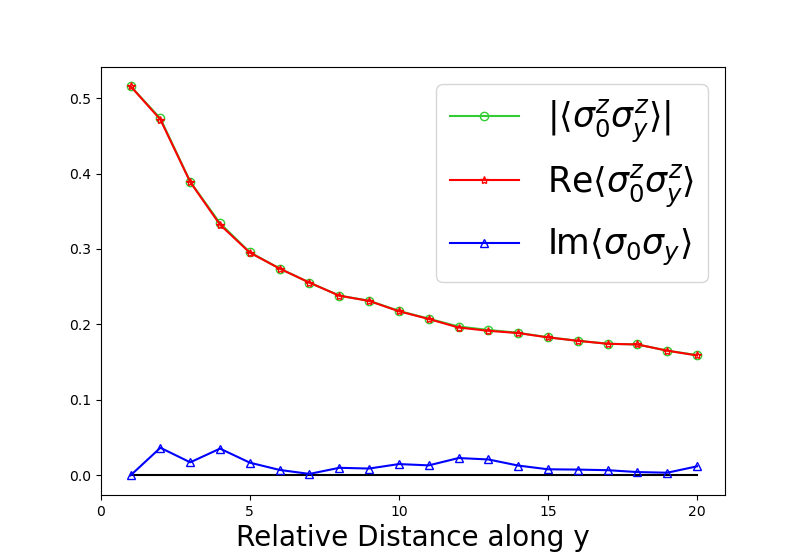}
\includegraphics[width=0.68\columnwidth]{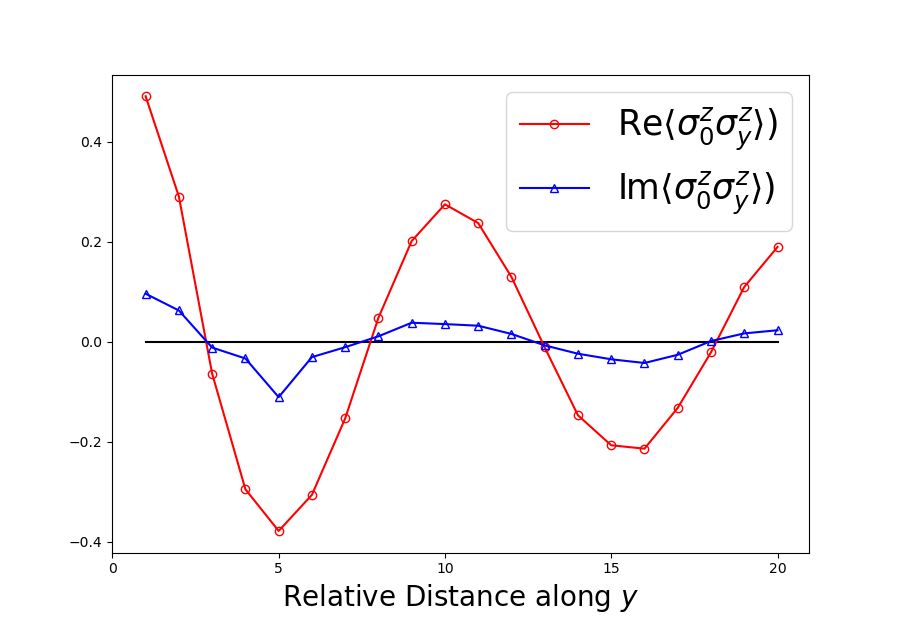}
\includegraphics[width=0.68\columnwidth]{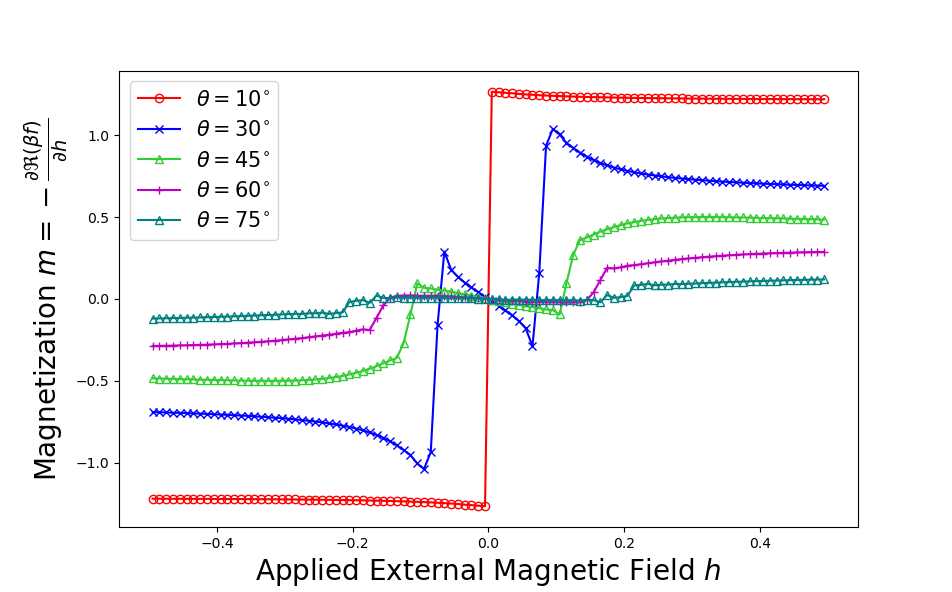}
\end{center}
\caption{Spin-spin correlations in the NFM1 and NFM2 phases, obtained via tensor renormalization group (TRG) algorithm.  Left-hand 
panels:  The spin-spin correlation function in the $x$ (time) direction (upper panel) and in the $y$ (intra-register) direction (lower 
panel) in the NFM1 phase, at the point $\tanh\beta = e^{i\pi/6}$.  Middle panels: The same, but in the NFM2 phase, at the point 
$\tanh\beta = 1.3 e^{i \pi/90}$.  Notice that the oscillations in the NFM2 phase run in the $y$ (intra-register) direction, while 
those in the NFM1 phase run in the $x$ (time) direction.  Right-hand upper panel: The spin-spin correlation function in the $x$ 
(time) direction in the NFM1 phase, at the point $\tanh\beta = e^{i\pi/3}$, for three different values of the applied longitudinal
 magnetic field.  Right-hand lower panel: The magnetization $m = -\pz\,\Rep(\beta f)/\pz h$ as a function of applied external
 magnetic field $h$, at various points $\tanh\beta = e^{i\theta}$.  The point $\theta = 10^\circ$ is in the FM phase; the others 
 are in the NFM1 phase.}
\label{f:correlations30}
\end{figure*}

Our TRG-computed correlation functions in the NFM1 phase are shown in the left-hand panels of Fig.~\ref{f:correlations30}.  
We observe modulated correlations along the direction of strong coupling ($x$ in the statistical mechanics setting; time in the 
quantum circuit picture). The correlations along the weak direction ($y$ in the statmech picture; inter-qubit in the quantum circuit
picture) are non-oscillatory and apparently power-law decaying.  In the next section we shall provide an interpretation of these 
results in terms of a fermionized version of the model. 
%
%

Our TRG results for the correlation functions in the NFM2 phase are shown in the middle panels of Fig.~\ref{f:correlations30}.  Surprisingly, we discover that the directions of the modulated and simply-decaying correlations are swapped in NFM2 relative to NFM1.  It is worth noting that for the NFM1 phase the $J_y \to 0$ limit is solvable and contains modulated correlations already, whereas the NFM2 phase does not exist in the decoupled limit, owing its existence to interchain interactions.

The upper right-hand panel of Fig.~\ref{f:correlations30} shows the correlation function in the $x$ (time) direction at a different point in the NFM1 phase, where the period of the temporal oscillations is shorter.  It also shows the effect on these correlations of the application of a longitudinal magnetic field.  We see that they survive almost entirely unaltered, \ie\ that this phenomenon is insensitive to the breaking of integrability.

The lower right-hand panel of Fig.~\ref{f:correlations30} shows the magnetization as a function of longitudinal field in the modulated phases. Up to some critical field (which depends on the extent of anisotropy), these modulated phases are stable. At this critical field a metamagnetic transition takes place and the system exits the modulated phase.

\begin{figure}[!b]
\begin{center}
\includegraphics[width=0.9\columnwidth]{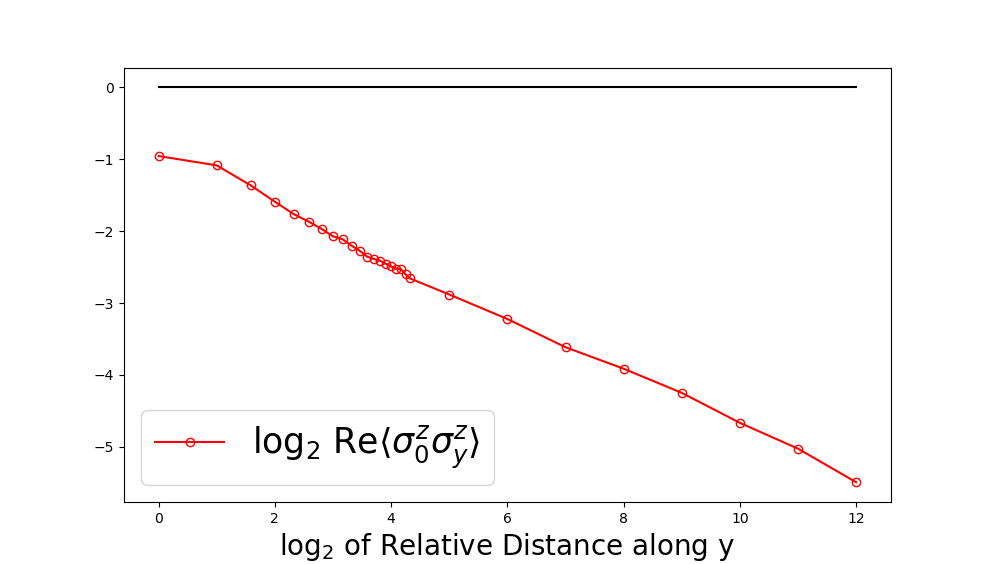}
\end{center}
\caption{Correlations along the non-oscillatory $y$ direction on a log-log scale, extracted from TRG for $\tanh \beta = e^{i \pi /6}$ and $J_y = 0.1 J_x$ (NFM1 phase), showing clear evidence of power-law decay with exponent $\sim 0.33$.}
\label{loglog}
\end{figure}

The TRG method makes it straightforward to compute correlations even for large systems, for particular 
exponentially spaced sets of distances. Evaluating the spatial decay of the correlator in the NFM1 phase, we see clear signs of algebraic decay (Fig.~\ref{loglog}), exactly as one would expect in a one-dimensional critical phase. 

\subsection{Fermionization and the origin of oscillations}\label{JWsec}

Finally, we present an approach to the zero-field problem that uses the Jordan-Wigner transformation to represent the spins/qubits in terms of fermionic degrees of freedom.  We shall show that the occurrence of correlations is due to a certain type of resonance between two eigenvalues of these fermionic operators.  We may use this picture to predict both the temporal period of the oscillations in the NFM1 phase and the spatial period of the oscillations in the NFM2 phase; in both cases, we find good agreement with our TRG results presented above.

In the Jordan-Wigner representation, we write the Pauli matrices on site $j$ as follows:
\begin{equation}
\begin{split}
X\ns_j & =  2 c\yd_j c\nd_j-1\\
Z\ns_j & =  (c\yd_j + c\nd_j)\,\prod_{l=1}^{j-1} e^{i\pi c\yd_l c\nd_l} \quad,
\label{JWX}
\end{split}
\end{equation}
where the operator $c_j$ annihilates a (spinless) fermion on site $j$ of the qubit register.  It follows that
\begin{equation}
Z\nd_j Z\ns_{j+1} = c\yd_j c\nd_{j+1} + c\yd_{j+1} c\nd_j - c\nd_j c\nd_{j+1} - c\yd_{j+1} c\yd_j \quad.
\label{JWZZ}
\end{equation}
We now Fourier transform the fermion operators with respect to the index $j$, \ie\ we move to a plane-wave basis in our qubit register:
\begin{equation}
c_ j= \frac{1}{\sqrt{M}}{\sum_k}' e^{-ikj}c_k\quad,\quad 
c\yd_j =\frac{1}{\sqrt{M}}{\sum_k}' e^{+ikj} c^\dagger_k\quad,
\label{cFT}
\end{equation}
where the prime on the sum restricts $k$ to the first Brillouin zone, \ie\ $k\in[-\pi,\pi)$.
In terms of these plane-wave operators, the components of our single-step time-evolution operator become
\begin{align}
\CW & =  \exp\!\bigg(2J\, {\sum_{k>0}}' \Big[\cos k \,\big(c\yd_k c\nd_k+c\yd_{-k} c\nd_{-k}\big) \\
&\hskip 1.5in + i \sin k\, \big(c\nd_{-k} c\nd_k - c\yd_k c\yd_{-k}\big)\Big]\bigg)\nonumber\\
\CV & = \exp\!\bigg(2\gamma\, {\sum_{k>0}}' \Big [c\yd_k c\nd_k+c\yd_{-k} c\nd_{-k} - 1 + \gamma^{-1}\ln A
\Big]\bigg)\quad,\nonumber
\end{align}
with $\CT=\CV^{1/2}\,\CW\,\CV^{1/2}$.

We may streamline our notation using Anderson pseudospin operators $\tau^\alpha_k$, defined as follows:
\begin{equation}
\tau_k^\alpha\equiv \begin{pmatrix} c\yd_k & c\nd_{-k} \end{pmatrix} \sigma^\alpha \begin{pmatrix} c\nd_k \\
c\yd_{-k} \end{pmatrix}\quad,
\label{e:pseudospins}
\end{equation}
where $\alpha\in\{0,1,2,3\}$.  In terms of these operators, the gates can now be re-written as follows:
\begin{align}
\CW & = {\prod_{k>0}}' \exp\!\Big(2 J\, \big(\tau^z_k \cos k + \tau^y_k \sin k\big) \Big) \label{eq:ZZgate_expressions}\\
\CV & = {\prod_{k>0}}' A^2 \exp\!\big(2 \gamma \tau^z_k\big)\label{eq:Xgate_expressions}\\
\CV^{1/2} &= {\prod_{k>0}}' A \exp\!\big(\gamma \tau^z_k\big)\quad.\label{eq:sqrtXgate_expressions}
\end{align}

The partition function $\CZ=\Tra\big(\CT^L\big)$ may be expressed as the product
\begin{align}
\CZ &= {\prod_{k>0}}' \Tra \big[\Theta_k^L\big] \label{e:fermZ}\\
\Theta\ns_k&\equiv A^2 \exp(\gamma \tau^z_k) \exp\!\big[2 J (\tau^z_k \cos k + \tau^y_k \sin k) \big] 
\exp(\gamma \tau^z_k)\quad.\nonumber
\end{align}
For each wavevector $k>0$, the operator $\Theta\ns_k$ has two eigenvalues, $\lambda_{k, \pm}$  
(see appendix \ref{app:fermionization}),
\begin{equation}
\lambda\ns_\pm(k)= 2\big(h\ns_k\pm\delta\ns_k\big)\quad,
\end{equation}
where $h\ns_k$ and $\delta\ns_k$ are given by
\begin{equation}
h\ns_k = \cosh (2 \beta ) \cosh (2J)+\sinh (2J)\cos (k) 
\end{equation}
and
\begin{align}
\delta\ns_k & =  \Big[\sinh^2 (2\beta)  \sinh^2(2J) \sin^2\! k \\
&\hskip 0.2in + \big[\cosh(2J)+\cosh(2\beta) \sinh(2J) \cos k \big]^2\,\Big]^{1/2}\quad,\nonumber
\end{align}
where we have used a mixed notation, trading coupling constants $\gamma$ and $A$ for $\beta$ -- recall the relations
$\tanh\gamma=e^{-2\beta}$ and $A=\sqrt{\coth\gamma-\tanh\gamma}$.  To analyze the late-time properties of the 
evolution, we find the largest-amplitude eigenvalue of $\Theta\ns_k$ (eq.\ref{e:fermZ}) for each wavenumber $k$. We denote 
this as  $\lambda\ns_0(k)$ and the corresponding right eigenvector as $\sket{\psi\ns_0(k)}$.  These eigenvalues determine 
the decay rate and precession of a typical initial condition at late times ($t\to \infty$):
\begin{equation}
\qket{0}\to {\prod_{k>0}}' \big[\lambda\ns_0(k)\big]^t\, \sket{\psi\ns_0(k)}\quad.
\end{equation}

We now discuss the behavior of this late-time state in terms of properties of the fermionic spectrum. Outside the NFM1 phase, 
either the $+$ or $-$ branch is consistently larger-amplitude throughout the Brillouin zone, and there is generically a unique 
steady state $\sket{0}$. In the NFM1 phase, however, the branches ``invert'' as a function of $k$, which is to say
$|\lambda\ns_+(k)| > |\lambda\ns_-(k)|$ for $k \approx 0$ but the opposite inequality holds for $k \approx \pm \pi$. 
At special momenta $\pm k^*$, the two eigenvalues are degenerate. Therefore, in the subspace of $\pm k^*$, the system 
never reaches a unique steady state, and instead one has persistent oscillations at the frequency 
\begin{equation}
\omega^* = \big|\arg\lambda\ns_+(k^*)- \arg\lambda\ns_-(k^*)\,\big|\quad.
\end{equation}
The band inversion point $k^*$ sweeps across the Brillouin zone as one progresses through the NFM1 phase, leading to incommensurate temporal modulations of varying frequency.

By comparing with the numerical solutions we see that this phase exhibits \emph{temporal} oscillations but no apparent spatial oscillations. While the momentum $2 k^*$ appears to be special in some sense, from the above argument, there is no simple relation between spectral degeneracies of the sort described above and spatial oscillations. To capture modulated correlations in the NFM2 phase, it is convenient instead to fermionize the model \emph{sideways}, by performing the Jordan-Wigner transformation along the $x$ axis (which hitherto we took to be the temporal direction). The ``band inversion'' described above now happens in the NFM2 phase, leading to oscillations in the \emph{spatial} direction (\ie, along $y$). 
\begin{table}
\begin{center}
\caption{Comparison of numerically estimated periods of order parameter oscillations ($T_{\rm TRG}$)  against exact 
Jordan-Wigner fermion periods, $T_{\rm JW}\equiv4 \pi/\omega^*$. Note the addtional factor of 2 due to a two-site unit-cell 
in the time-direction implicit in the definition of the one step evolution operator. First three rows corresponds to points inside 
NFM1, where modulations are along the $x$-axis, while the last row is in NFM2.}
\includegraphics[width=\columnwidth]{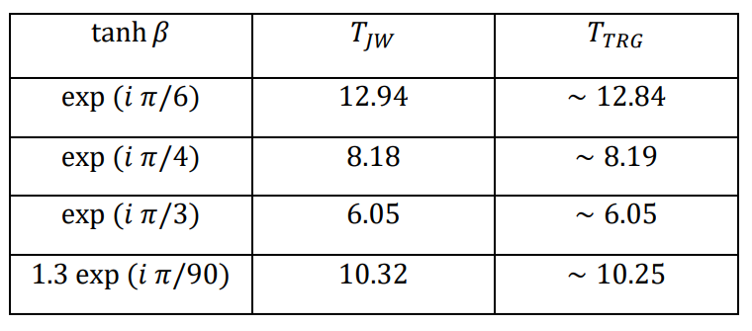}
\end{center}
\end{table}

\section{Discussion}

In the present work we have exploited the correspondence between non-unitary quantum circuits and complex temperature 
statistical mechanics to construct a simple quantum circuit that has a surprisingly rich phase diagram, including a phase with
incommensurate temporal order. Such incommensurate time crystals do not seem to occur in closed systems; nor do they occur in
 one-dimensional open quantum systems with short-range interactions, for entropic reasons. Our results suggest that an important 
 class of quantum circuits that exhibit incommensurate time-crystalline order are spacetime duals of circuits that realize 
 incommensurate density-wave phases. In this simple one-dimensional case, such phases occur only for complex couplings, 
 but in more general settings it might be possible to write down quantum circuits that cool the system into a ground state with 
 density-wave order~\cite{stoudenmire2010minimally}. These would also be spacetime dual to temporally modulated phases. 

In practice, post-selection is an expensive operation requiring effort that scales exponentially in the area of the quantum circuit. 
Thus, practical realizations of the physics discussed here will be restricted to circuits that are either very shallow or involve only a 
small number of qubits evolved for a long time. These map onto Ising ladders at complex temperature, which can be solved using 
the methods discussed above (App.~\ref{app:ladders}). We find that signatures of the modulated phases are present even for 
systems with modest numbers of spins ($M = 5$), which should be realistic to explore in a variety of present-day experiments.

\section{Acknowledgments} \label{s:ack}
We are grateful to F. Beichert, R. Moessner, F. Pollmann, D. Huse, E. Fradkin, S. Kivelson, W. Bialek, Alexander Abanov, 
M. Stoudenmire, Tzu-Chieh Wei, N. Pomata, John Cardy and Michael Fisher for stimulating discussions over the years.  
CAH gratefully acknowledges financial support from UKRI under grant number EP/R031924/1; he is also grateful to Rice 
University for a visiting appointment in spring 2019, where part of this work was completed. SB and VO acknowledge support 
from the NSF DMR Grant No. 1508538 and US-Israel BSF Grant No. 2014265.

\appendix
\section{Complex-temperature statistical mechanics of Ising ladders}
\label{app:ladders}
In this appendix, we review previously obtained results on loci of Fisher zeros as they coalesce into branch cuts for $M>1$.  We also present some new results where we identify the regions of the complex-temperature plane in which the correlation lengths exceed ten lattice spacings.  These latter results foreshadow the form of the phase diagram in the $M \to \infty$ limit, Fig.~\ref{f:phasediag}.

Quite generally, the only place where long-range order can occur in an infinite Ising system is on the same contours where the partition function zeros for the finite system concentrate.  This is because the condition for the two is the same, viz.\ that the largest two eigenvalues of the transfer matrix become equimodular. As explicitly demonstrated analytically for $M=1$ and numerically for $M>1$, Fisher zeros merge into $\approx M$ branch cuts in the complex temperature plane in the thermodynamic (long-time) limit, thus allowing for a smooth evolution of correlation functions, with the correlation time becoming infinite on branch cuts and the correlations themselves retaining an oscillatory character inherited from the relative phase of the two dominant eigenvalues\cite{Beichert2013}.

What happens to the spin-spin correlation length (decay time) in the regions between the contours?  The answer is that, between the $M$ contours that are `coalescing' into the ordered region in the 2D Ising model, the correlation length stays very high.  The existence of ferromagnetism over a finite region of the complex-temperature plane may be anticipated by noticing that phases of the largest and second-largest eigenvalues `lock' to each other (as $M\to\infty$).  This appears to be the only type of correlated `phase' that occurs in the isotropic case.  Anisotropic lattices, however, appear to support another type of `gapless' correlated phase, which exhibits multiple long length-scales, and which shows precursor signatures in the behavior of the correlation lengths in the finite-$M$ case.

\begin{figure}
\begin{center}
\includegraphics[width=0.95\columnwidth]{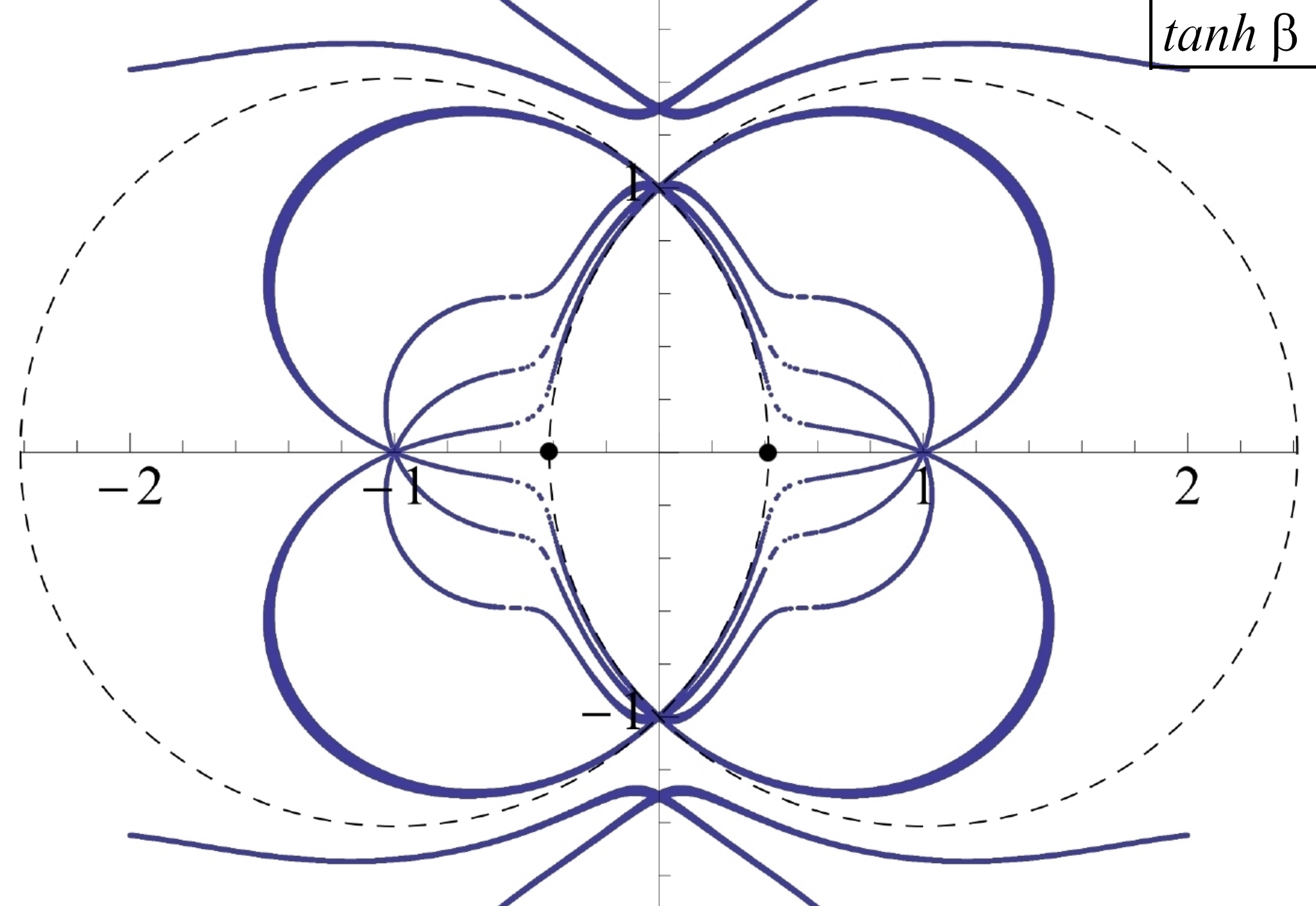}
\end{center}
\caption{(Reproduced from F.~Beichert {\it et al.}\cite{Beichert2013}) The contours containing the zeros of the partition function in the complex temperature plane for Ising spin ladders with 4 legs and isotropic spin-spin interactions ($J_x = J_y$).  Dashed lines represent the expected location of Fisher zeros of the isotropic 2D Ising model\cite{Fisher1965}. The total number of contours is equal to the number of legs; their location was found to be only weakly sensitive to boundary conditions in the short (rung) direction and insensitive to the boundary conditions in the long (infinite) direction.}
\label{f:laddersfeli}
\end{figure}

We begin by reviewing our prior results\cite{Beichert2013}, in which the limiting behavior of Fisher zeros in ladders was computed --- see Fig.~\ref{f:laddersfeli}.  It can be shown, via a low-temperature expansion, that the number of contours (branch cuts) emerging from the two zero-temperature points $\tanh \beta=\pm 1$ is equal to $M$.  It is less clear how to compute the total number of contours, although contours that do not emerge from $\tanh \beta = \pm 1$ do not appear to proliferate, and may be strongly dependent on boundary conditions.

Fisher's original proposal overlooked this behavior entirely.  The exact solutions of the Ising model by Onsager and several others in the years that followed\cite{anisotropic2d} usually consist of several contributions, only one of which dominates in the thermodynamic limit.  Fisher's original argument for generalizing Yang-Lee results was based on a seemingly incorrect procedure whereby he analytically continued only the portion of the result that was important at real temperature.  As explicitly demonstrated by Beichert {\it et al.\/}, this produces entirely wrong patterns of zeros in ladders. Remarkably, however, Fisher's approximate solution is accurately reproduced by the unbiased TRG computational scheme applied to the 2D lattice.

\begin{figure}
\begin{center}
$\qquad$ $M=2$, longest $\qquad \qquad$
$M=2$, second-longest
\includegraphics[width=0.9\columnwidth]{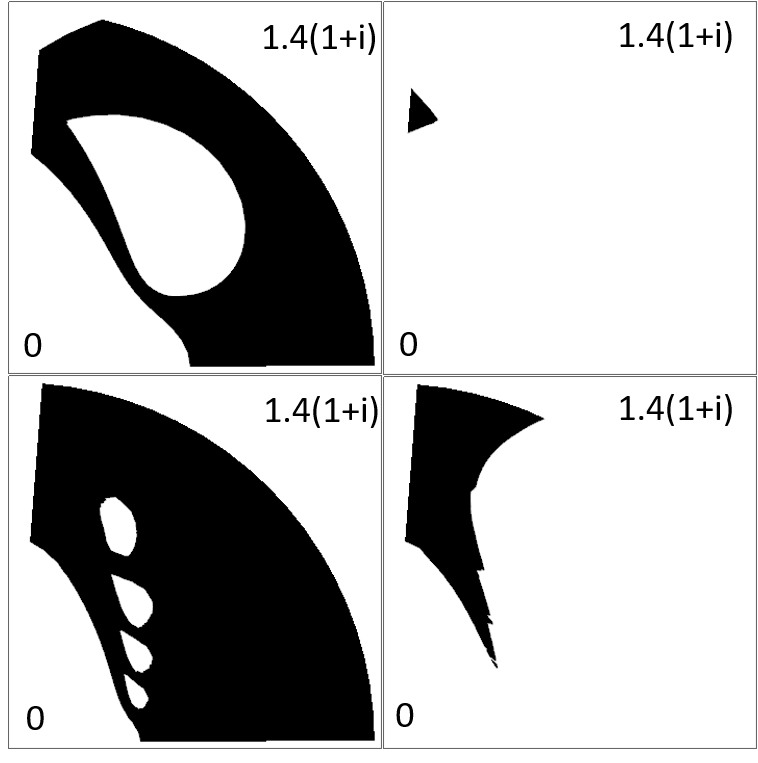}

\vspace{-1mm}
$\qquad$ $M=5$, longest $\qquad \qquad$
$M=5$, second-longest
\end{center}
\caption{The top-right quadrant of the complex $\tanh \beta$ plane of Fig.~\ref{f:laddersfeli} for an $M$-leg Ising spin ladder with isotropic interactions, $J_x = J_y = 1$.  Upper panels:\ for a 2-leg ladder, the longest correlation length (left) and the second-longest (right), both in the `long' direction, \ie\ along the legs of the ladder.  Lower panels:\ the same for a 5-leg ladder.  The regions in which the correlation length in question exceeds 10 lattice sites are shaded black.  Note that, except near the unitary point $\tanh \beta = i$, only the longest correlation length shows significant structure.}
\label{f:ladders}
\end{figure}

Next we examine the growth of correlation lengths in the $x$ (time) direction as we increase $M$.  Each correlation length is controlled by the ratio between one of the subdominant eigenvalues of the transfer matrix, $\lambda_j$ ($j>0$), and the dominant one, $\lambda_0$.  In the left-hand panels of Fig.~\ref{f:ladders}, the shaded areas mark the regions of the complex-temperature plane in which the longest correlation length, \ie\ the one controlled by $\lambda_1/\lambda_0$, is greater than ten lattice spacings.  This is shown for the two-qubit case (top left) and for the five-qubit case (bottom left).  It is clear that our arbitrarily determined threshold of 10 lattice sites is already exceeded for $M=5$ in the entire crescent region outlined by Fisher's original 2D proposal.  The right-hand panels show the same information but for the second-longest correlation length, \ie\ the one controlled by $\lambda_2/\lambda_0$.  We note that there is no other long correlation length present; as we can see from these plots, the eigenvalue $\lambda_2$ does not approach $\lambda_0$ except near the unitary point $\tanh \beta = i$, \ie\ the ferromagnetic phase is `gapped'.

\begin{figure}
\begin{center}
$\qquad$ $M=2$, longest $\qquad \qquad$
$M=2$, second-longest
\includegraphics[width=0.9\columnwidth]{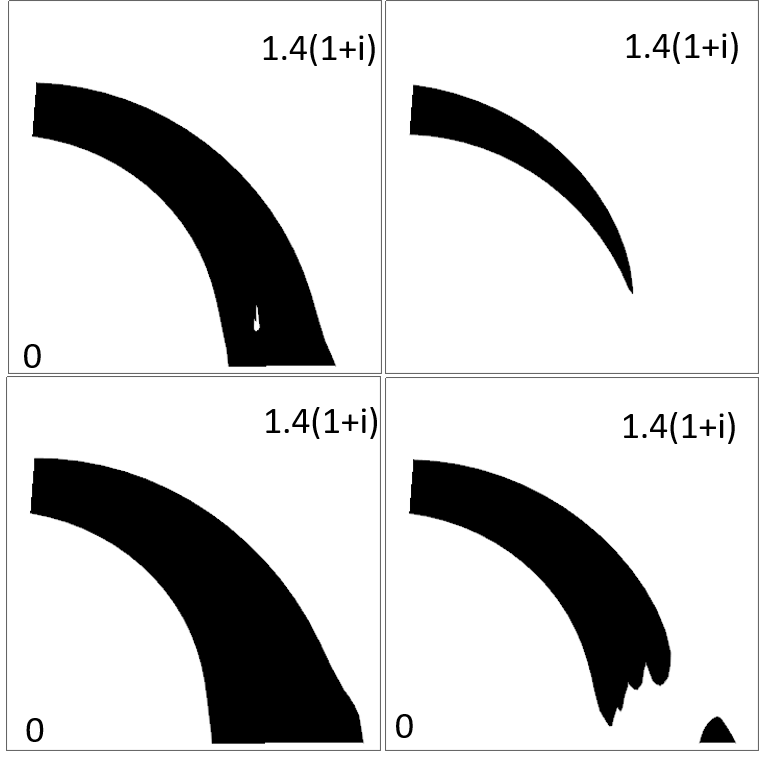}

\vspace{-1.8mm}
$\qquad$ $M=5$, longest $\qquad \qquad$
$M=5$, second-longest
\end{center}
\caption{The same as Fig.~\ref{f:ladders}, but for the case of anisotropic interactions, $J_y = J_x/10 = 0.1$.  Note that, in a region roughly corresponding to the `NFM1' and `NFM2' regions of Fig.~\ref{f:phasediag}, the second-longest correlation length also shows non-trivial structure.}
\label{f:laddersaniso}
\end{figure}

We next turn to the case of anisotropic couplings, $J_y < 1$.  It was observed by van Saarloos and Kurtze \cite{Saarloos1984} that, 
in this case,  Fisher's approximation to the partition function produces highly complex patterns of zeros.  For simple integer fractions
$J_y=1/n$ it is possible to compute and plot the contours\cite{BasuUnpub} and observe an erratic pattern that does not show simple 
convergence to the limit of decoupled qubits.  Numerically exact computations for finite-$M$ transfer matrices, however, produce 
a nicely regular and convergent pattern, from which Fig. \ref{f:laddersaniso} was obtained.  The ordered phase is reduced in its 
extent as we have reduced $J_y$ tenfold.  As with the isotropic model we observe a growth of the longest correlation length as 
we increase $M$.  However, in this anisotropic case we also see additional regions away from the unitary point in which the 
second-longest correlation length also becomes long:\ one tracking the unit circle, and another small patch on the real 
$\tanh \beta$ axes at values exceeding 1.  These locations are suggestively similar in shape and location to the NFM1 and 
NFM2 phases in Fig.~\ref{f:phasediag}.

\section{TRG results for the magnetization : isotropic case}
\begin{figure}
\begin{center}
\includegraphics[width=\columnwidth]{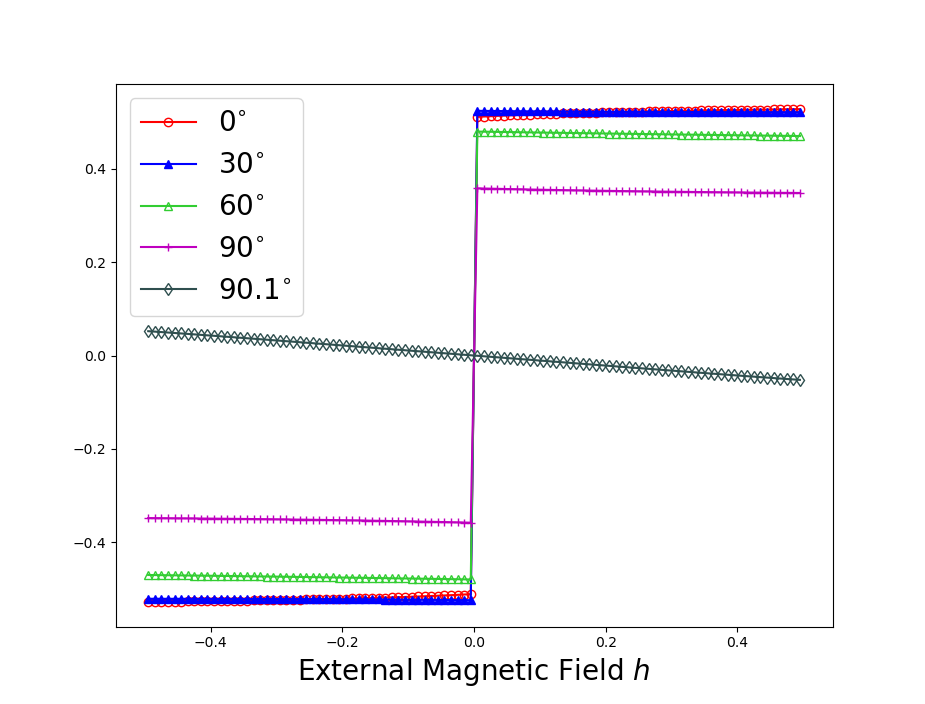}
\includegraphics[width=\columnwidth]
{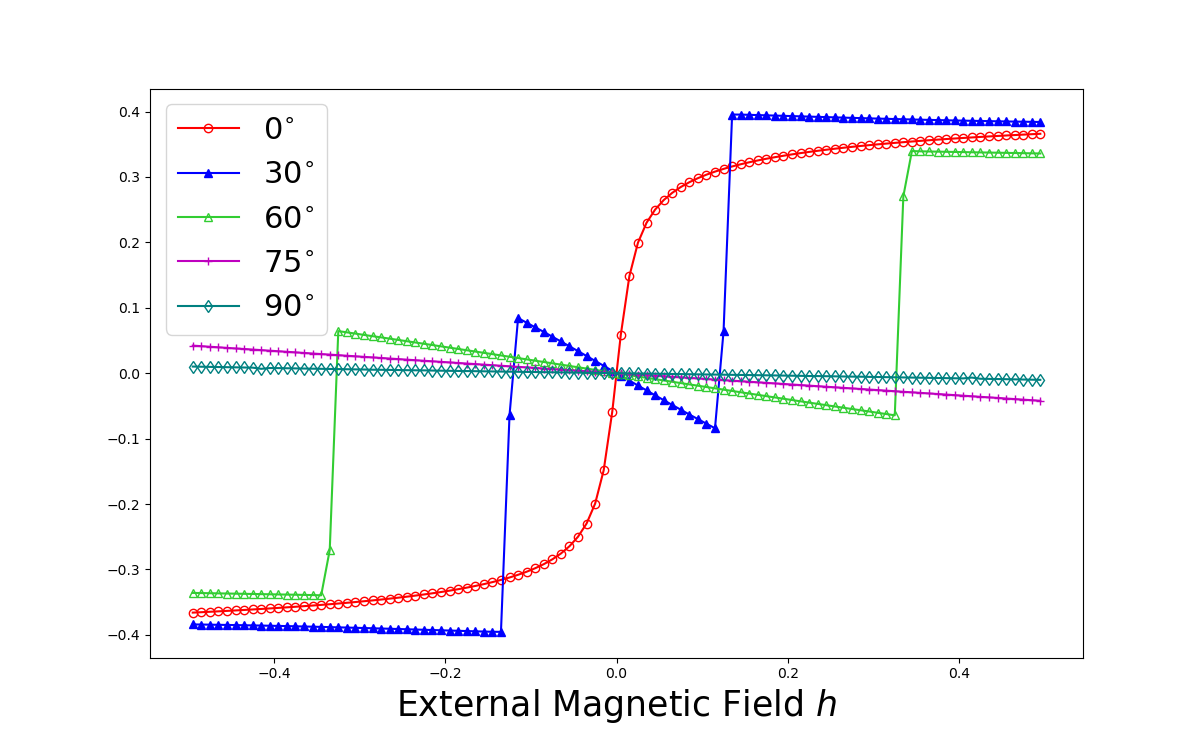}
\end{center}
\caption{Magnetization, $m = -\pz\,\Rep(\beta f)/\pz h$ as a function of applied uniform magnetic field in the two phases of the isotropic 2D Ising model in the complex $\sinh(2 \beta)$ plane.  Upper panel:\ ferromagnetic phase, $|\sinh(2 \beta)| = 1.3$. Lower panel:\ metamagnetic response in the paramagnetic phase, $|\sinh (2 \beta)|=0.9$.}
\label{f:magnetization}
\end{figure}
In this appendix we present our TRG results for the behavior of the isotropic Ising model in a longitudinal field.

Based on the patterns of Fisher zeros, it seems that the isotropic Ising model in zero applied magnetic field exhibits, for general 
complex temperature, a first-order transition between a paramagnetic (PM) and a ferromagnetic (FM) phase.  We explicitly verify 
this through a direct computation of the magnetization process in the vicinity of that phase boundary.

The phase boundary between the PM and FM phases forms the famous double crescent in the $\tanh \beta$ plane, but becomes a 
yet simpler unit circle in the complex $\sinh 2 \beta$ plane:\ this appears to be a more natural variable for the isotropic case.  
A trajectory with a fixed modulus of $\sinh 2\beta$ slightly above/below 1 allows us to study the evolution of the character of 
the transition as we move away from the real-temperature critical point.

In Fig.~\ref{f:magnetization}, we show our results for the magnetization as a function of field for a variety of angles along two 
such circular contours in the complex $\sinh 2 \beta$ plane.  In the upper panel, we observe that the FM phase indeed exhibits
spontaneous magnetization, the magnitude of which drops to zero as the point $\arg \sinh 2\beta=\pi/2$ is approached.  The 
PM phase shows initially linear response, followed by a metamagnetic jump, as expected near conventional first-order transitions\cite{chaikin1995principles}.  Interestingly, there is a large domain of \emph{diamagnetic} response, which would 
be a rather peculiar state of affairs for a more conventional statistical problem.

\section{Fermionization}
\label{app:fermionization}
In this appendix, we provide the details of the various steps of our fermionization approach to the zero-field case.

In \S \ref{JWsec} it was shown that the partition function $\CZ=\Tra\big(\CV^{1/2}\,\CW\,\CV^{1/2}\big)^{\!L}$ is given by
$\CZ= \prod_{k>0}' \Tra \big[\Theta_k^L\big]$, where
\begin{equation}
\Theta\ns_k\equiv A^2\, e^{\gamma \tau^z_k}\, e^{2 J (\tau^z_k \cos k + \tau^y_k \sin k) }\,e^{\gamma \tau^z_k}\quad.
\end{equation}
For our purposes here, considering one $k$ mode at a time, we write $\Btau=\{X,Y,Z\}$ for the Pauli matrices
 $\big\{\tau^x_k,\tau^y_k,\tau^z_k\big\}$.  We then have
\begin{equation}
e^{\gamma Z} = \cosh\gamma + Z\sinh\gamma
\end{equation}
and
\begin{align}
e^{2J (Y\!\sin k + Z\cos k)}&=\cosh(2J) \\
&\hskip 0.4in + \sinh(2J) \,(Y\!\sin k + Z\cos k)\quad.\nonumber
\end{align}
We find $\Theta\ns_k=2\big[d\ns_0(k) +\Bd(k)\cdot\Btau\big]$ with
\begin{equation} 
\begin{split}
d\ns_0(k)&=\cosh(2\beta)\cosh(2J) + \sinh(2J) \cos k\\
d\ns_x(k)&=0\\
d\ns_y(k) &=\sinh(2\beta)\sinh(2J) \sin k\\
d\ns_z(k) &= \cosh(2J)+ \cosh(2\beta)\sinh(2J)\cos k\quad.
\end{split}
\end{equation}
Thus the eigenvalues of $\Theta\ns_k$ are given by
\begin{equation}
\begin{split}
\lambda\ns_\pm(k) &= 2 d\ns_0(k) \pm 2\sqrt{d_y^2(k)+d_z^2(k)}\\
&=2 d\ns_0(k) \, \Big\{1 \pm \sqrt{\Omega_+(k)\,\Omega_-(k)}\Big\}
\end{split}
\end{equation}
where
\begin{equation}
\Omega\ns_\pm(k)\equiv {d\ns_z(k) \pm i d\ns_y(k)\over d\ns_0(k)}\quad.
\end{equation}

\noindent{\sl Resonances} : The {\it resonance condition\/} $\big|\lambda\ns_+(k)\big|=\big|\lambda\ns_-(k)\big|$ 
thus pertains when
\begin{equation}
\Omega_+(k)\,\Omega_-(k) =-\alpha^2 \in \mathbb{R}_-\quad.
\end{equation}
This entails
\begin{equation}
\begin{split}
0&=d_+(k)\,d_-(k) +\alpha^2 d_0^2(k)\\
&=(1+\alpha^2)\cosh^2(2\beta)\cosh^2(2J)-\sinh^2(2\beta)\\
&\qquad\ +(1+\alpha^2)\cosh(2\beta)\sinh(4J)\cos k\\
&\qquad\qquad +(1+\alpha^2)\sinh^2(2J)\cos^2\!k\quad, 
\end{split}
\end{equation}
which is a quadratic equation in $\cos k$, with the solution
\begin{equation}
\cos k = {1\over\sinh(2J)}\Bigg\{\!\!-\cosh(2\beta)\cosh(2J) \pm {\sinh(2\beta)\over\sqrt{1 + \alpha^2}}\Bigg\}
\label{eq:Res3}
\end{equation}
Now we know that $\alpha \in \MR$ and hence the closed form expression for the resonance condition can be derived from equating $\Imp(\alpha^2)=0$ and $\Rep(\alpha^2)\geq 0$.

From (\ref{eq:Res3}), we can derive the expression for $\alpha^2$ as
\begin{equation}
\alpha^2 = {\sinh^2(2\beta)\over [\cosh(2\beta)\cosh(2J) + \cos k \sinh(2J)]^2}-1
\label{eq:Res4}
\end{equation}
and the subsequent closed form resonance conditions for persistent oscillations immediately follow
\begin{align}
 \Imp\!\left({\sinh^2(2\beta)\over[\cosh(2\beta)\cosh(2J) + \cos k \sinh(2J)]^2}\right) &= 0 
\label{eq:Res5}\\
\Rep\!\left({\sinh^2(2\beta)\over [\cosh(2\beta)\cosh(2J) + \cos k \sinh(2J)]^2}\right) &\geq 1\ .
\label{eq:Res6}
\end{align}
Note that both conditions in (\ref{eq:Res5}) and (\ref{eq:Res6}) need to be satisfied simultaneously in order to 
get the correct phase diagram, which matches the phase diagram obtained from TRG and analytically continued Onsager solution 
(see Fig.~\ref{f:phasediag}). 

\smallskip

\noindent {\sl Steady state} -- We now obtain an expression for the steady state, where each $(k,-k)$ mode pair is in an
eigenstate of $\Theta\ns_k$.  With $\Omega\ns_\mu(k)\equiv d\ns_\mu(k)/d\ns_0(k)$ for $\mu\in\{x,y,z\}$, we have
\begin{equation}
\Theta\ns_k=2d\ns_0(k)\begin{pmatrix} 1 + \Omega\ns_z(k) & -i\Omega\ns_y(k) \\
i\Omega\ns_y(k) & 1-\Omega\ns_z(k)\end{pmatrix}\quad.
\end{equation}
As $\Theta\ns_k$ is in general non-Hermitian, its right and left eigenvectors are not necessarily related by complex
conjugation, and are given by
\begin{equation}
\begin{split}
\sket{R\ns_\pm(k)} &= \begin{pmatrix} \mu\ns_\pm(k) \\ \mu\ns_\mp(k) \end{pmatrix}\\
\sbra{L\ns_\pm(k)} &= \pm\CN\ns_k\begin{pmatrix} \mu\ns_\pm(k) &,& -\mu\ns_\mp(k) \end{pmatrix} \quad,
\end{split}
\end{equation}
where
\begin{equation}
\mu\ns_\pm(k) = \sqrt{\Omega\ns_+(k)}\pm\sqrt{\Omega\ns_-(k)} 
\end{equation}
and $\CN\ns_k=1/4\sqrt{\Omega\ns_+(k)\,\Omega\ns_-(k)}$.  These states are normalized so that
$\sbraket{L\ns_a(k)}{R\ns_b(k)}=\delta\ns_{ab}$ (with no complex conjugation of the left eigenvector).

In the $t\to\infty$ limit, and at each wavevector $k\in(0,\pi)$, one of these states is selected -- the one
corresponding to the greater value of $\big|\lambda\ns_\pm(k)\big|$.  The surviving state's wavefunction is
given by the appropriately normalized right eigenvector, and the asymptotic state is thus of the BCS form,
\begin{equation}
\qket{\Psi(t\to\infty)}={\prod_{k>0}}' \CC\ns_k\,\Big[\mu\ns_\mp(k) + \mu\ns_\pm(k)\,c\yd_k c\yd_{-k}\Big] \,
\qket{0}\quad,
\end{equation}
where $\CC_k = 1/\sqrt{2\big(|\Omega_{k,+}| + |\Omega_{k,-}|\big)}$, and where $\qket{0}$ is the 
Fock space vacuum, equivalent to the state $\qket{\!\!\dar\dar\cdots\dar}$ for all the $k$-space Anderson pseudospins.

%


\end{document}